\documentstyle[floats,aps,aipbook,psfig]{revtex}
\righthead{Nemiroff et al.}
\lefthead{Time-Dilation, Log N - Log P, and Cosmology}
\begin{document}
\title{Time-Dilation, Log N - Log P, and Cosmology}
\author{R.J. Nemiroff,$^{1,2}$
        J.P. Norris,$^1$
        J.T. Bonnell,$^3$ and J.D. Scargle$^4$}
\address{$^1$NASA/Goddard Space Flight Center, Greenbelt, MD 20771\\
         $^2$George Mason University, Fairfax, VA 22030\\
         $^3$Universities Space Research Association\\
         $^4$NASA/Ames Research Center, Moffett Field, CA 94035}

\maketitle

\begin{abstract}
We investigate whether a simple cosmology can fit GRB results in both time
dilation and Log N - Log P simultaneously.  Simplifying assumptions
include: all GRBs are spectrally identical to BATSE trigger 143, $\Omega=1$
universe, and no luminosity and number density evolution.  Observational
data used includes: the BATSE 3B peak brightness distribution (64-ms time
scale), the Pioneer Venus Orbiter (PVO) brightness distribution, and the
Norris et al. time dilation results for peak aligned profiles presented at
this meeting. We find acceptable cosmological fits to the brightness
distributions when placing BATSE trigger 143 at a redshift of 0.15 $\pm$
0.10. This translates into a $(1 + z_{dim}) / (1 + z_{bright})$ factor of
about 1.50 $\pm$ 0.50 between selected brightness extremes of the Norris et
al. sample. Norris et al. estimate, however, that $(1 + z_{dim}) / (1 +
z_{bright})$ $\approx$ 2.0 $\pm$ 0.5 when considering duration tests. The
difference is marginal and could be accounted for by evolution. We
therefore find that evolution of GRBs is preferred but not demanded. 
\end{abstract}

\section*{Introduction}

There are two main results evident with BATSE data which, when taken
together, appear consistent with a cosmological origin \cite{{BP}}
\cite{{TP}}.  The first is the isotropic nature of the BATSE GRB
distribution, and the second is the apparently confined nature of the peak
brightness distribution \cite{{GJF}} \cite{{MEE}}. Cosmological fits to the
brightness distribution have been made by several research teams
\cite{{CDD}} \cite{{TW}} \cite{{EH}}. An independent third result is now
claimed to be consistent with a cosmological origin of GRBs: time dilation
\cite{{JPNtd1}} \cite{{RJNspec}} \cite{{JPNtd2}} \cite{{CP}}
\cite{{Mallozzi}}. In this paper we investigate whether the most recent time
dilation results of Norris et al. \cite{{JPNhunt95}} and Bonnell et al.
\cite{{Bonnell}} are consistent with the 3B peak brightness
distribution. Fenimore \& Bloom \cite{{FAB}} imply that the two might not
be consistent with a single cosmological setting, as the redshifts they
derived from the Norris et al. \cite{{JPNtd1}} time dilations are much
greater than those find implied by the brightness distribution and
uniformity assumptions. 
 
Data from the BATSE 3B catalog was accessed from the Compton GRO Science
Support Center.   BATSE bursts with T90 durations greater than 2 seconds
and peak fluxes greater than 1.827 photons cm$^{-2}$ sec$^{-1}$ were
included. BATSE is at least 95 \% complete to this peak flux, according to
the published BATSE 2B trigger efficiency table. 

\begin{figure*}
\leavevmode
\psfig{file=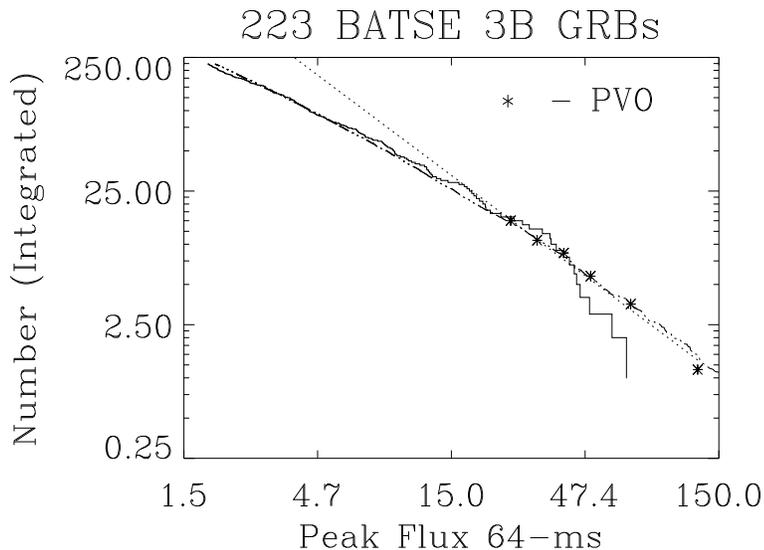,height=3.0in,width=4.0in}
\caption{ The combined BATSE - PVO brightness distribution fit to an
$\Omega=1$ cosmology.  The solid histogram steps show BATSE 3B data. The
asterisks show PVO data normalized to the BATSE rate. The dotted line
represents a canonical uniform distribution with -1.5 slope line. The
dot-dashed line shows a Monte Carlo fit to an $\Omega=1$ cosmology of
standard candle GRBs with no evolution.  The best fit had BATSE trigger 143
placed at a redshift of 0.15, and statistically acceptable fits placed 143
within a redshift of 0.1 of this value. } 
\end{figure*} 

PVO brightness distribution data was taken from Table 1 of Fenimore and
Bloom \cite{{FAB}}. The conversion from PVO rate to equivalent BATSE rate
was found by demanding that the rates be equivalent at the dimmest complete
PVO peak flux bin. The conversion factor of 1.25 between peak flux in the
PVO energy band 100 - 500 keV and peak flux in the BATSE energy band 50 -
300 keV was given by Fenimore in a private communication. 

To test for consistency we created Monte Carlo simulations.  We first
generated theoretical log N - log P (here P stands for peak flux in photons
cm$^{-2}$ sec$^{-1}$ on the 64-ms time scale) by considering BATSE trigger
143 a burst with a canonical spectrum. BATSE trigger 143 is one of the
brightest and best studied GRBs: its spectrum is well known because of good
counting statistics and because it was also seen by the EGRET and COMPTEL
instruments on board CGRO; its time series is well studied because of good
counting statistics and because it was only the 10th cosmic GRB detected by
BATSE.  We use only the spectrum at the peak - more specifically for 1
second centered on the bin of peak counts. 

We quantified 143's photon spectrum and generated a theoretical cosmology
by throwing it randomly in an $\Omega = 1$ universe.  In the subsequent
Monte-Carlo simulation, we then re-measured each new ``143" burst being
careful to consider all cosmological dimming, time-dilating, and reddening
factors on the actual 143 spectrum and burst rate.  No evolution was
considered, however. 

\begin{figure*} \leavevmode
\psfig{file=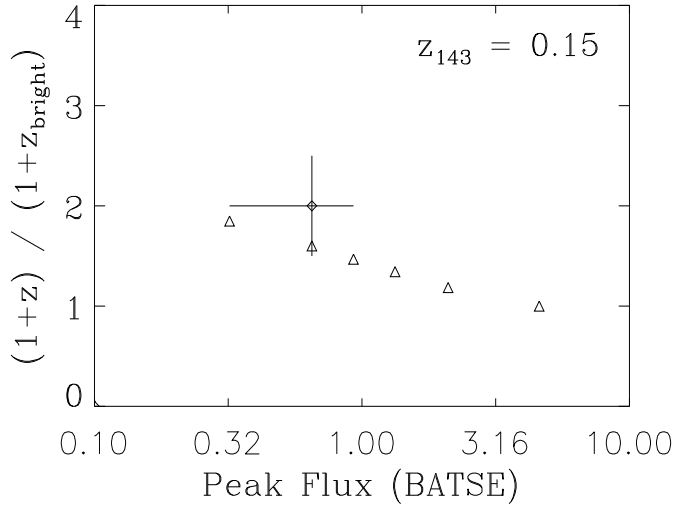,height=3.0in,width=4.0in} 
\caption{Comparison of the time-dilation results implied by the brightness
distribution plots with the time-dilation results measured by Norris et al.
for a simple cosmological model.  The triangles represent the expected
time-dilation from fits to the Log N - Log P for the simple cosmology
described and for BATSE trigger 143 placed at a redshift of 0.15. The
diamond with the error bars is the time-dilation result extracted from
Norris et al. measured by the peak alignment technique.} 
\end{figure*} 

We then numerically compared the theoretical brightness distributions
generated with a combined BATSE - PVO brightness distribution. The
comparison was done using the two-distribution KS test.  The test is
sensitive to not just the shape of the two distributions but their dynamic
range in peak flux. The only parameter varied to maximize the goodness of
fit was the actual redshift of the measured 143 spectrum itself. 

\section*{Results}

First, ignoring time dilation, we found it quite possible to fit the
combined BATSE - PVO brightness distribution with this simple cosmological
paradigm.  Such a fit is shown in Figure 1.   In this fit BATSE trigger 143
was placed at a redshift of 0.15.  Acceptable fits at the 1 $\sigma$ level
were found for redshifts between 0.05 and 0.25. 

Next, we wished to test whether these fits were consistent with the most
recent Norris et al. \cite{{JPNhunt95}} measurement of time-dilation.
Figure 2 shows effective time dilations from the best fit to the cosmology
implied by Figure 1, along with the recent time dilation result for the
peak alignment method \cite{{JPNhunt95}}. Only one redshift-corrected
time-dilation point can be used to date. 

Figure 2 shows that the time-dilation factor between the brightness groups
in question is about 2.0, while the time-dilation factor implied by the
brightness distribution fit in the above cosmology is about 1.5.  There is
some uncertainty in the Norris et al. \cite{{JPNhunt95}} factor however,
which is about 0.50 in redshift based on the quoted uncertainties and the
jitter of the time dilation points.  This places the two time-dilation
factors about one $\sigma$ apart. One may note that the time dilation in
this bin is only about two $\sigma$ away from no time-dilation at all - but
the inference that time-dilation itself has only marginal statistical
significance would be deceiving. This is because neighboring unrelated data
points also stray in the same sense from the line, so that the cumulative
probability of all the points straying is greater \cite{{JPNtd2}}
\cite{{JPNhunt95}}. As the full time-dilation redshift-decomposition has
only been estimated for a single point as yet, we can only estimate that
the statistical probability would be at least a factor of a few higher,
were points from all brightness classes accounted for. 

We also note that were a $1 - \sigma$ higher fit redshift used for 143 in
the brightness distribution, a much better agreement would be found between
the time-dilation factor implied by the brightness distribution and that
measured by the Norris et al. \cite{{JPNhunt95}} time dilation analysis. 

\section*{Conclusions} 

In general we find the cosmologies implied by the Log N - Log P and the
time-dilation are only in good agreement for some methods of time-dilation
determination. For peak alignment and auto-correlation measures there is a
disagreement in the sense that the burst redshifts implied by the Norris et
al. \cite{{JPNtd1}} \cite{{JPNtd2}} \cite{{JPNhunt95}} time dilations are
greater than those implied by the combined BATSE - PVO brightness
distributions. Since the measured errors are large, however, even this
simple non-evolutionary cosmology is not rigorously excluded - to greater
than about 2 $\sigma$ - from explaining both results simultaneously. {\it
We therefore conclude that evolution is preferred but not demanded.}


\begin{references}
\bibitem{BP}B. Paczynski, Nature {\bf 355}, 521 (1992).
\bibitem{TP}T. Piran, ApJ {\bf 389}, 45 (1992).
\bibitem{GJF}G.J. Fishman et al., ApJ Supp. {\bf 92}, 229 (1994).
\bibitem{MEE}C.A. Meegan et al., these proceedings.
\bibitem{CDD}C.D. Dermer, PhysRevL {\bf 68}, 1799 (1992).
\bibitem{TW}W.A.D.T. Wickramasinghe, R.J. Nemiroff, J.P. Norris, C.
Kouveliotou, G.J. Fishman, C.A. Meegan, R.B. Wilson, \& W.S. Paciesas, ApJ
{\bf 411}, L55 (1993). 
\bibitem{EH}A.G. Emslie  \& H.M. Horack,  ApJ {\bf 435}, 16 (1994).
\bibitem{JPNtd1}J.P. Norris et al., ApJ {\bf 424}, 540 (1994).
\bibitem{RJNspec}R.J. Nemiroff, J.P. Norris, J.T., Bonnell,
W.A.D.T. Wickramasinghe, J.D.  Scargle, C. Kouveliotou, 
W.S. Paciesas, G.J. Fishman, \& C.A. Meegan, ApJ {\bf 435}, L133 (1994).
\bibitem{JPNtd2}J.P. Norris et al., ApJ {\bf 439}, 542 (1995).
\bibitem{CP}E. Cohen \& T. Piran, ApJ {\bf 444}, 25 (1995).
\bibitem{Mallozzi}R.S. Mallozzi et al., ApJ {\bf 454}, 597 (1995).
\bibitem{JPNhunt95}J.P. Norris et al., these proceedings.
\bibitem{Bonnell}J.T. Bonnell et al., these proceedings.
\bibitem{FAB}E.E. Fenimore and J.S. Bloom, ApJ {\bf 453}, 25 (1995).

\end{references}
\end{document}